\documentclass[sigconf,screen]{acmart}

\usepackage{graphicx} %
\usepackage{geometry}
\usepackage{array}
\usepackage{enumitem}
\usepackage[acronym]{glossaries}
\makeglossaries
\usepackage{multirow, makecell}
\usepackage{tikz}
\usepackage{adjustbox}
\usepackage{booktabs}
\usepackage{diagbox} 
\usetikzlibrary{arrows.meta, positioning, shapes.geometric}
\usepackage{tabularx}
\usepackage{booktabs}
\usepackage{threeparttable}
\newcolumntype{Y}{>{\centering\arraybackslash}X} 
\usepackage{caption}

\newacronym{pgm}{PGM}{probabilistic graphical model}

\newacronym{nn}{NN}{neural network}
\newacronym{gbdt}{GDBT}{Gradient Boosted Decision Trees}
\newacronym{llm}{LLM}{large language model}
\newacronym{serp}{SERP}{search engine results page}
\newacronym{rctr}{RCTR}{rank-based CTR model}
\newacronym{dctr}{DCTR}{document-based CTR model}
\newacronym{rcm}{RCM}{random click model}
\newacronym{pbm}{PBM}{position-based model}
\newacronym{cm}{CM}{cascade model}
\newacronym{ubm}{UBM}{user browsing model}
\newacronym{dcm}{DCM}{dependent click model}
\newacronym{dbn}{DBN}{dynamic bayesian network model}
\newacronym{ncm}{NCM}{neural click model}
\newacronym{csm}{CSM}{click sequence model}
\newacronym{pscm}{PSCM}{partially sequential click model}
\newacronym{tcm}{TCM}{temporal click model}
\newacronym{thcm}{THCM}{temporal hidden click model}
\newacronym{emalgorithm}{EM}{expectation-maximization}
\newacronym{ips}{IPS}{inverse propensity scoring}
\newacronym{mle}{MLE}{maximum likelihood estimation}
\newacronym{ctr}{CTR}{click-through rate}
\newacronym{ccm1}{CCM}{carousel click model}
\newacronym{xpa}{XPA}{cross-positional attentions}
\newacronym{gubm}{GUBM}{grid-based user browsing model}
\newacronym{ccm2}{CCM}{click chain model}
\newacronym{cbcm}{CBCM}{comparison-based click model}
\newacronym{rbnn}{RBNN}{rank-biased neural network model}
\newacronym{drlc}{DRLC}{debiased reinforcement learning click model}
\newacronym{cacm}{CACM}{context-aware click model}
\newacronym{graphcm}{GraphCM}{graph-enhanced click model}
\newacronym{aicm}{AICM}{adversarial imitation click model}
\newacronym{fscm}{FSCM}{f-shape click model}
\newacronym{tacm}{TACM}{time-aware click model }
\newacronym{pctm}{PCTM}{probability click tracking model }
\newacronym{bbm}{BBM}{bayesian browsing model}
\newacronym{pcc}{PCC}{post-click click model}
\newacronym{gcm}{GCM}{general click model}
\newacronym{bss}{BSS}{bayesian sequential state model}
\newacronym{fetcm}{FE-TCM}{filter-enhanced transformer click model}

\allowdisplaybreaks

\newtheorem{assumption}{Assumption}

\author{Jingwei Kang}
\affiliation{%
  \institution{University of Amsterdam}
  \city{Amsterdam}
  \country{The~Netherlands}
}
\email{j.kang@uva.nl}
\orcid{0009-0003-9283-4060}

\author{Maarten de Rijke}
\orcid{0000-0002-1086-0202}
\affiliation{%
  \institution{University of Amsterdam}
  \city{Amsterdam}
  \country{The~Netherlands}
}
\email{m.derijke@uva.nl}

\author{Harrie Oosterhuis}
\orcid{0000-0002-0458-9233}
\affiliation{%
  \institution{University of Amsterdam}
  \city{Amsterdam}
  \country{The~Netherlands}
}
\email{h.r.oosterhuis@uva.nl}

\copyrightyear{2026}
\acmYear{2026}
\setcopyright{cc}
\setcctype{by}
\acmConference[SIGIR '26]{Proceedings of the 49th International ACM SIGIR Conference on Research and Development in Information Retrieval}{July 20--24, 2026}{Melbourne, VIC, Australia}
\acmBooktitle{Proceedings of the 49th International ACM SIGIR Conference on Research and Development in Information Retrieval (SIGIR '26), July 20--24, 2026, Melbourne, VIC, Australia}
\acmDOI{10.1145/3805712.3809657}
\acmISBN{979-8-4007-2599-9/2026/07}



\title{Following the Eye-Tracking Evidence: Established Web-Search Assumptions Fail in Carousel Interfaces}
\begin{document}

\begin{abstract}
Carousel interfaces have been the de-facto standard for streaming media services for over a decade.
Yet, there has been very little research into user behavior with such interfaces, which thus remains poorly understood.
Due to this lack of empirical research, previous work has assumed that behaviors established in single-list web-search interfaces, such as the F-pattern and the examination hypothesis, also apply to carousel interfaces, for instance when designing click models or evaluation metrics.
We analyze a recently-released interaction and examination dataset resulting from an eye-tracking study performed on carousel interfaces to verify whether these assumptions actually hold.

We find that 
(i)~the F-pattern holds only for vertical examination and not for horizontal swiping; additionally, we discover that, when conditioned on a click, user examination follows an L-pattern unique to carousel interfaces; 
(ii)~click-through-rates conditioned on examination indicate that the well-known examination hypothesis does not hold in carousel interfaces; 
and (iii)~contrary to the assumptions of previous work, users generally ignore carousel headings and focus directly on the content items.
Our findings show that many user behavior assumptions, especially concerning examination patterns, do not transfer from web search interfaces to carousel recommendation settings.
Our work shows that the field lacks a reliable foundation on which to build models of user behavior with these interfaces.
Consequently, a re-evaluation of existing metrics and click models for carousel interfaces may be warranted.
\end{abstract}

\maketitle

\glsresetall

\section{Introduction}
\label{sec:introduction}

With the widespread adoption of carousel interfaces by major streaming platforms such as Netflix and Spotify, these interfaces have become the de-facto standard for streaming media recommendations.
They present recommendations in vertically stacked rows, where each row is a single carousel: a horizontally swipeable list of items organized around a clear theme; together, such rows form the \textit{carousel interface}~\cite{10.1145/3383313.3412217, 10.1145/2959100.2959174, 10.1145/3604915.3610638}, as shown in Figure~\ref{fig:recgaze}.
Despite their ubiquitous real-world application, there is very little research on user behavior with carousel interfaces~\cite{10.1145/3708359.3712130, 10.1145/3604915.3610638, 10.1145/3726302.3730301, 10.1145/3742413.3789166}, especially when compared to the large amount of work on other interfaces~\cite{10.1145/1008992.1009079, 10.1145/1076034.1076063,doi:10.1177/1541931214581234, 10.1145/3077136.3080799, LORIGO20061123}.
As a result of this gap, our current understanding of how users interact with these recommendations is far from complete; and this knowledge gap significantly impedes the development of user models, evaluation metrics, and ranking algorithms tailored for carousel interfaces.

Another consequence of this gap is that the few existing studies that do address carousel interfaces have to rely on behavioral assumptions that lack empirical validation~\citep{10.1145/3511095.3531278, 10.1145/3643709, 10.1145/3452918.3465493, 10.1145/3450614.3461680, 10.3389/fdata.2022.910030}.
Some studies have introduced the two-dimensional NDCG (N2DCG or NDCG 2D) for carousel recommendation that adopt the \textit{golden triangle} (a.k.a.\ \textit{F-pattern}) assumption from single-list interfaces in the web search setting~\cite{10.1145/3452918.3465493, 10.3389/fdata.2022.910030}. 
These works assume that users explore a full-page carousel interface beginning at the top-left corner, with attention decreasing both horizontally (left to right) and vertically (top to bottom).
Recently, \citet{10.1145/3643709} proposed the first click model specifically for carousel interfaces.
Their approach assumes that users examine carousels sequentially from top to bottom, identify an attractive carousel solely based on its topic label without examining the items it contains, and examine individual items only when the corresponding carousel is attractive. 
The model further relies on the very well-known \textit{examination hypothesis}, assuming that a click occurs only if an item is both examined and attractive.
These assumptions are either transferred directly from single-list interfaces in web search, are adapted from them or are simplistic heuristics.
Consequently, the validity of models and metrics built on these assumptions remains uncertain. 
This critical gap leads to the central question of this work:
\begin{quote}
\textbf{\textit{Are these legacy assumptions about user behavior accurate in the carousel setting, and if not, what does real user behavior with carousel interfaces actually look like?}}
\end{quote}
To address this research gap, we perform a detailed analysis based on a recently-released dataset from a dedicated eye-tracking study for carousel interfaces~\citep{10.1145/3726302.3730301}.
By systematically comparing the behavioral assumptions underlying current models with observations of empirical user data, we found no strong evidence to support their similarities; instead, we found significant differences between the assumptions and the data.

Our main contributions are:
\begin{itemize}[leftmargin=*,nosep]
    \item We introduce a \textbf{\textit{L-shaped}} examination pattern, revealed when conditioning on clicks.
    We further demonstrate that the traditional F-pattern does not hold for horizontal swiping.
    Thus, we have two findings that highlight unique aspects of examination behavior with carousel interfaces.
    \item We re-evaluate the examination hypothesis by estimating the conditional probability of clicking given examination. Our empirical observations heavily imply the standard examination hypothesis does not apply to carousel interfaces.
    \item We challenge vertical navigation assumptions by showing that users largely ignore carousel topic labels and go directly to their content, contradicting the navigation logic assumed by existing carousel click models.
\end{itemize}

\noindent%
Our findings reveal an urgent need to rethink the behavioral assumptions for carousel recommendations and provide valuable insights for building more realistic, behavior-based models and metrics that can truly capture user experiences.

\section{Related Work}

In modern information retrieval (IR) research, the analysis of user behaviors  (e.g., click logs \cite{10.1145/1526709.1526711, 10.1145/1341531.1341545}, dwell times \cite{10.1145/2556195.2556220, 10.1145/2600428.2609468, 10.1145/3077136.3080799}, and gaze data \cite{10.1145/1076034.1076063, 10.1145/1008992.1009079, 10.1145/3077136.3080799}) has greatly informed our understanding of how users examine and interact with search results, which in turn has shaped many of the core assumptions underlying widely used click models \cite{10.1145/1341531.1341545,10.1145/1390334.1390392,10.1145/1498759.1498818,10.1145/1526709.1526711, 10.1145/3209978.3209990} and evaluation metrics \cite{10.1145/3308558.3313514, 10.1145/345508.345545, 10.1145/582415.582418}.

\subsection{User behavior in non-carousel interfaces}

\subsubsection{Single-list interface}
\label{sec:relatedwork:singlelist}
\begin{figure}[t]
\centering
\includegraphics[width=\linewidth]{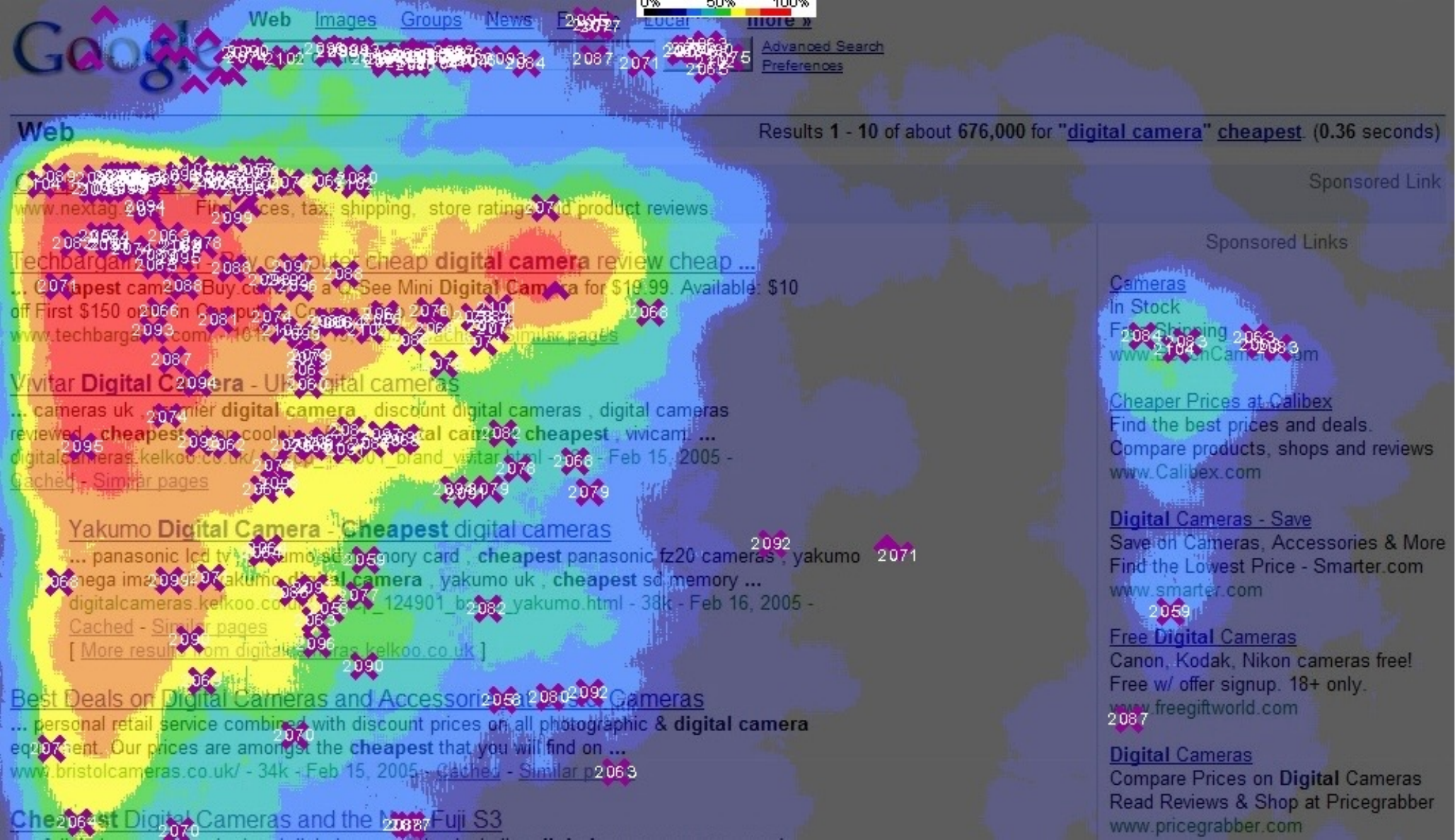}
\caption{Visual attention heatmap of a Google results page released by Enquiro, EyeTools and Did-It.}
\label{fig:golden triangle}
\end{figure}

The single-list interface is the most classical and extensively studied interface in IR \cite{10.1145/1498759.1498818,10.1145/1526709.1526711, chuklin_markov_rijke_click_models_2015,10.1145/1008992.1009079}. 
A large body of user behavior analysis based on eye-tracking and click logs has revealed a consistent examination pattern in single-list settings, known as the \emph{golden triangle}, sometimes also referred to as the F-pattern \cite{Granka2008}.
It is characterized by a triangle or F-shape that indicates users concentrate heavily on the top-left area of the page, i.e., see Figure~\ref{fig:golden triangle}.

Moreover, eye tracking studies have motivated the \emph{examination hypothesis}, which assumes that a document must be examined before being clicked and that click probabilities are a product of examination and attractiveness probabilities~\cite{10.1145/1242572.1242643}. 
Based on this hypothesis, concrete click models introduce different additional assumptions from observed user behaviors. 
For example, inspired by the strong click decay across position from top to bottom \cite{10.1145/1076034.1076063,10.1145/1008992.1009079}, some models assume a sequential examination process in which users examine results from top to bottom \cite{10.1145/1341531.1341545,10.1145/1390334.1390392,10.1145/1498759.1498818,10.1145/1526709.1526711}. In contrast, motivated by findings of non-sequential examination behaviors, such as revisiting previously examined results \cite{LORIGO20061123}, other models have been developed to account for these behaviors \cite{10.1145/2124295.2124334,10.1145/2766462.2767712}.

Similar to click models, mainstream ranking evaluation metrics are also motivated by user examination behavior. 
For example, DCG~\cite{10.1145/345508.345545} and NDCG~\cite{10.1145/582415.582418} adopt logarithmic discounting over ranks to reflect the top-down examination pattern.

\subsubsection{Grid interface}
User behavior in grid-based interfaces has been studied across multiple grid-style layouts with different spatial organizations~\cite{10.1145/1743666.1743736,doi:10.1080/10447318.2013.846790,10.1145/2959100.2959150, 10.1145/3077136.3080799, 10.1145/1935826.1935873}. 
However, unlike  single-list interfaces, existing studies show that user behavior in grid interfaces is highly dependent on the specific task setting and interface design, leading to different attention and interaction patterns.

On the one hand, studies of regular grid layouts, where items are neatly arranged in rows and columns, with a fixed number of items per row, show attention patterns that are very similar to those observed in single-list interfaces.
For example, \citet{10.1145/2959100.2959150} observed that gaze behavior in regular grid interfaces often follows an F-shaped pattern, with user attention concentrated in the top-left corner rather than at the center of the screen. Similarly, \citet{10.1145/1935826.1935873} illustrated the probability distribution of user clicks across regular grid layouts, showing a primary golden triangle of attention in the top-left corner, while also revealing a secondary silver triangle in the bottom-right area.

On the other hand, \citet{10.1145/3077136.3080799} studied user examination behavior in grid-based image search with irregular layouts, where image size is different and the number of images per row is not fixed. 
In this setting, they found a clear middle-position bias, rather than the traditional F-shaped pattern. 
Motivated by this specific behavior, a grid-based interaction model~\citep{10.1145/3209978.3209990} and an evaluation metric~\citep{10.1145/3308558.3313514} were proposed.

\subsection{User behavior in carousel interfaces}

In contrast to single-list and grid interfaces, there is very little research on carousel interfaces~\cite{10.3389/fdata.2023.1239705,10.1145/3383313.3412217,10.1145/3450614.3461680,10.1145/3604915.3610638} and even less research on direct user behavior regarding carousel interfaces.
Currently, only a single published eye-tracking study exists that specifically focuses on carousel interfaces~\cite{10.1145/3726302.3730301}, providing preliminary results on how user visual attention is distributed in the carousel layout.
Additionally, this study released the RecGaze dataset that contains their collected eye-tracking data.
Presently, there is only one paper that has used this dataset: its analysis specifically studies the gaze transition between carousels and between items in carousels~\cite{10.1145/3742413.3789166}.

Due to this scarcity of observational studies, user behavior in carousel interfaces remains poorly understood, making it unclear how click models or evaluation metrics should be developed for this setting. %
To address this gap, existing studies have relied on assumptions of user behavior as a basis for developing models and metrics for the carousel setting:
For example, two recent works~\cite{10.1145/3452918.3465493,10.3389/fdata.2022.910030} have proposed N2DCG for carousel recommendation by directly adopting the golden triangle (or F-pattern) assumption from traditional web search. 
In addition, \citet{10.1145/3511095.3531278} assume that users first identify an attractive carousel based on its topic label before considering items; based on these assumptions, they further propose the \gls{ccm1}~\cite{10.1145/3643709}.
However, to the best of our knowledge, none of these assumptions have been empirically verified in carousel interfaces.

\section{Assumptions Based on Existing Literature}
\label{sec:expect}

As stated in Section~\ref{sec:introduction}, we aim to verify whether assumptions adopted from non-carousel interface settings, i.e., single-list or grid-based layouts, are accurate for carousel interfaces.
This section outlines assumptions about user behavior in carousel interfaces that have been proposed based on previous studies in other settings and have been adopted by existing work.
Subsequently, the following sections compare these assumptions with the RecGaze dataset~\cite{10.1145/3726302.3730301} to evaluate whether they are actually accurate, and thereby, verify whether they should be used to construct assumptions for carousel interfaces.

\subsection{Assumptions 1 \& 2: User  examination patterns in carousel interfaces}\label{sec:expectation 1+2}

For web search interfaces, it has been widely observed that user attention is generally focused on the top-left corner of the page; this phenomenon is often referred to as the \emph{golden triangle} or \emph{F-shaped pattern} (see Section~\ref{sec:relatedwork:singlelist} and Figure~\ref{fig:golden triangle}). 
This attention pattern has been consistently observed in both traditional single-list \cite{Granka2008} and grid-based layouts \cite{10.1145/2959100.2959150, 10.1145/1935826.1935873}.
Reasonably, one could expect a similar pattern in carousel interfaces:
\begin{assumption}\label{exp:expectation 1}
    \textbf{User examination (unconditioned) follows an F-shaped pattern}, where most attention is given to the most top-left item and attention consistenly decays in the vertical (top to bottom) and horizontal directions (left to right).
\end{assumption}

\noindent%
We note that existing work that introduced the NDCG2D and N2DCG metrics~\cite{10.1145/3452918.3465493, 10.3389/fdata.2022.910030} are grounded in this assumption.

Another common assumption about carousel interfaces is that user examination follows an \emph{hierarchical and sequential} browsing process. This was first formally proposed by \citet{10.1145/3511095.3531278, 10.1145/3643709}.
This assumption comes from the design of carousel interfaces.
As the carousel design specifically aims to conveniently presents items with a common topic, and browsing can be done both vertically between carousels and horizontally within carousels.
In addition, a common assumptions in single-list interfaces is that user examination follows a cascade model~\cite{10.1145/1341531.1341545,10.1145/1390334.1390392,10.1145/1498759.1498818,10.1145/1526709.1526711}: results are observed sequentially until an attractive result is found.
Taking these assumptions together, it seems natural to expect user examination follows the intended design in a cascading manner:
\begin{assumption}\label{exp:expectation 2}
    \textbf{User examination conditioned on a click follows a hierarchical pattern}, where the carousel topics above a clicked item and the items to the left of the clicked item have been examined.
    Items to the right of a clicked item or in carousels below a clicked item are not expected to be examined; whilst items above the carousel with the click could be examined.
\end{assumption}

\noindent%
We note that in their \gls{ccm1}~\cite{10.1145/3643709}, \citet{10.1145/3511095.3531278, 10.1145/3643709} actually make a stronger assumption since they also assume that any carousels preceding the one with the click should either be examined completely or have no item examinations at all.
Since it is stronger, refuting Assumption~\ref{exp:expectation 2} would thereby also refute their assumption.

Finally, we note that Assumption~\ref{exp:expectation 1}~\&~\ref{exp:expectation 2} do not contradict each other, since the hierarchical pattern in \emph{examination conditioned on clicks} can coincide with an F-pattern in \emph{unconditional examination}.

\subsection{Assumption 3: User clicks in carousels adhere to the examination hypothesis}\label{sec:expectation 3}
The \emph{examination hypothesis} was first proposed by \citet{10.1145/1242572.1242643} for the traditional single-list web search interface and has since been widely adopted by various click models~\cite{chuklin_markov_rijke_click_models_2015, 10.1145/1341531.1341545, 10.1145/1390334.1390392, 10.1145/1498759.1498818, 10.1145/1526709.1526712}.
This hypothesis assumes that a document must be examined to be clicked, and that the click event $C$ is conditioned on a latent variable examination $E$, such that the probability of a click is factorized as:
\begin{equation}
P(C = 1) = P(E = 1) \cdot P(C = 1 \mid E = 1),
\end{equation}
which implies that $P(C = 1 \mid E = 0) = 0$.

The conditional probability of a click on an item given examination ($E=1$) is commonly assumed to be affected only by the item itself, interpreted as its relevance or attractiveness:
\begin{equation}
P(C = 1 \mid E = 1, \text{item}, \text{position}) =P(C = 1 \mid E = 1, \text{item}).
\end{equation}
Thus, once an item has been examined, the decision to click solely depends on its attractiveness (and not on its position or presentation).
Given the wide adoption of the examination hypothesis, it seems natural to expect it to hold for carousel interfaces as well, which implies the following assumption regarding click probabilities and examination:
\begin{assumption}\label{exp:expectation 3}
\textbf{Click probabilities of an item conditioned on examination are \textbf{equal} over all display positions.}
In other words, the conditional probability of a click on a specific item, given that the user has examined it, has the same value for all possible positions where it can be displayed.
\end{assumption}

\noindent%
We note that \gls{ccm1}, the only existing click model for carousel interfaces, assumes the examination hypothesis~\cite{10.1145/3643709}, but it has never been empirically verified for carousel interfaces.

\subsection{Assumption 4: User navigation relies on the carousel titles of carousel interfaces}\label{sec:expectation 4}

A key characteristic of carousel interface design is that each carousel (row) has a descriptive title or genre label above its upper left corner.
Each title provides a semantic summary of the content of its carousel and is designed to allow users to quickly find a list of items that is relevant to their current interest through vertical browsing, before spending effort on horizontal navigation through items~\citep{10.1145/3511095.3531278}.
Accordingly, one can expect specific behavior that is encouraged by this design:
\begin{assumption}\label{exp:expectation 4}
    \textbf{User examination considers the title of a carousel before examining any items within that carousel}, where it is possible that only the title of a carousel is examined and none of its items.
\end{assumption}

\noindent%
We note that \gls{ccm1}~\citep{10.1145/3643709} makes this assumption as well, as its click model explicitly describes users start with considering the topic of a carousel in order to decide whether to consider any items within the carousel.

\section{Carousel Eye-Tracking Data and Analysis}
\subsection{The RecGaze dataset}
\begin{figure}[t]
    \centering
    \includegraphics[width=\linewidth, alt={Netflix carousel interface}]{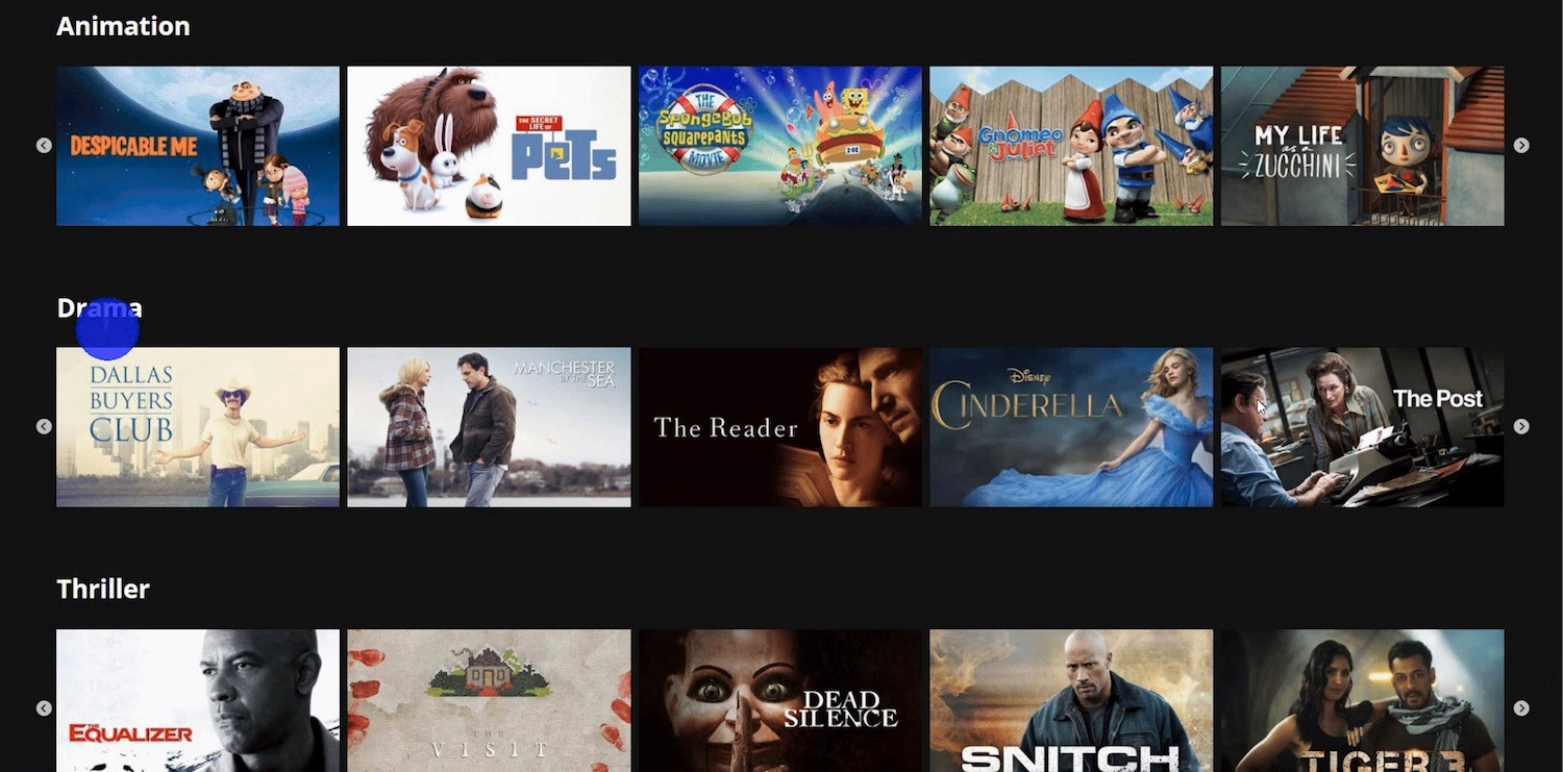}
    \caption{Design of a screen in the RecGaze eye-tracking study. The blue circle represent a user's current gaze point. }
    \label{fig:recgaze}
\end{figure}

For our analysis, we use the eye tracking data of the RecGaze dataset~\cite{10.1145/3726302.3730301}.\footnote{\url{https://zenodo.org/records/15270518}} To the best of our knowledge, this is the only publicly available eye tracking data on carousel interfaces.
This study had participants navigate a carousel interface that mimics those of major streaming platforms such as Netflix to select a movie to watch out of the presented recommendations.
Each screen displayed ten genre-based carousels, each containing fifteen movies which are divided into three \emph{pages} of five movies, arrows on the side of each carousel allow users to navigate between pages; see Figure~\ref{fig:recgaze} for an example.

Users were given one of three distinct tasks: free browsing, semi-free browsing, and search; for this work, we only consider the free browsing task as it most closely approximates natural user behavior in a realistic scenario.
For the free browsing task, users were instructed to click on a single movie that they want to watch per screen; after each click, they were presented with two questions regarding their familiarity with the selected movie and the reason for their choice; subsequently, the next screen was presented.
In total, 87 participants were involved in this study, each was presented 30 screens for the free browsing task.
After filtering out screens where participants indicated they incorrectly clicked on a movie (i.e., by accident), we are left with 2375 valid screens (with one click per screen) on which we base our analysis.

An important aspect for our study is how the screens were populated for the RecGaze dataset.
The carousels were displayed in completely random (vertical) order per screen; randomization took place during screen creation, not when displaying to users, meaning that each user was presented with an identical set of 30 screens.
Per carousel, the movies were displayed in order of popularity, where popularity was measured in the number of IMDb votes they received.\footnote{\url{https://imdb.com}}
To mitigate potential bias against older movies, only movies released between 2000 and 2023 were selected.
Additionally, to mitigate a potential bias against highly popular movies, the top 150 movies with the most votes on IMDb were excluded.
In summary, each screen displays ten genre-based carousels in random order, where each carousel shows fifteen movies of that genre in descending order of their popularity on IMDb.

Since examination is a latent variable that cannot be observed directly, we approximate user examination at each interface position using fixations.
Fixations shorter than 60 ms are already excluded by the dataset~\citep{10.1145/3726302.3730301}, as such brief fixations are unlikely to reflect active cognitive processing. 
Consistent with common practices in click model research \cite{10.1145/1341531.1341545,10.1145/1390334.1390392,10.1145/1498759.1498818,10.1145/1526709.1526711}, we model examination as a binary variable: a specific item (e.g., a movie thumbnail or carousel title) is considered \emph{examined} if at least one valid fixation falls within its area of interest (AOI); otherwise, it is treated as \emph{unexamined}.

\subsection{Machine-learning-based smoothing}
\label{sec: data smoothing}
Our analysis is mainly based around the observed click and examination frequencies in the dataset.
However, as one would expect in a carousel interface, items at the start of carousels (on the left side) attract substantially more user attention and interaction than subsequent items.
Hence, while examination data is available across all positions, with even tail items being examined in at least 250 screens, click data on items at the end of carousels are very rare.
In the extreme, there are even some positions that did not receive any clicks across all 2375 screens at all, meaning no conditional frequencies (conditioned on clicks) are observed at all.
This leads to a difficult comparison of frequencies based on highly different numbers of data points.
Thus, while we do report the observed frequencies, we apply an additional model-fitting step to estimate and predict probabilities which mitigates this problem with variance.

We fit a tree-based model (i.e., gradient boosted decision trees) to the interaction data such that it can predict $P(E_{i,j} = 1 \mid C_{i',j'} = 1)$, the probability that position $(i,j)$ is examined given that position $(i',j')$ is clicked, where $(i,j) \not= (i',j')$.
Here, $i$ denotes the carousel row and $j$ denotes the movie column within the carousel.
We provide the model with features characterizing both the target position $(i,j)$  and the conditioning position $(i',j')$. 
These include their absolute coordinates on the interface, as well as relative features capturing their spatial relationship, such as the distance between them.
The model thus assumes that examination probabilities vary smoothly 
as a function of spatial features, such that positions with similar 
spatial characteristics tend to have similar probabilities.
This is consistent with how user attention propagates in a carousel interface, where browsing behavior is inherently continuous. 
For the conditional examination probability, the dataset provides us with $2375 \times (160-1)$ examples.
Subsequently, the model can be used to provide a prediction for every possible combination of target position and conditioning position.
The resulting predictions are not prone to variance and are therefore smoother and have more regularity than the raw frequencies in the data, making it easier to reveal patterns and perform our analysis.

\begin{figure*}[t]
  \centering
  \includegraphics[width=\linewidth]{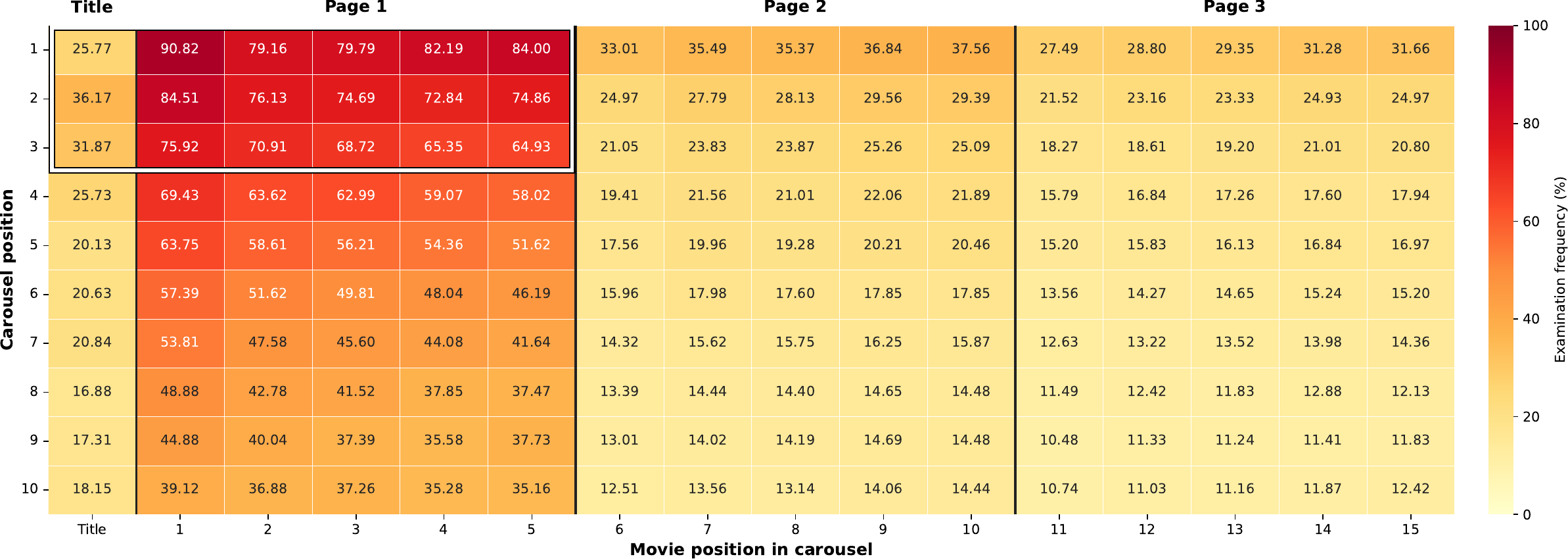}
  \caption{Empirical examination frequency heatmap across positions and pages. The number in each small unit represents the frequency at that position. The white box indicates the initially visible items and titles. Refuting Assumption~\ref{exp:expectation 1}:  We can observe a dual-focus F-shaped pattern on Page~1 and mirrored F-shaped patterns within Page~2 \& 3. Importantly, we also observe that there is no global F-pattern that goes across pages.
  Refuting Assumption~\ref{exp:expectation 4}: We also observe that most users do not examine the titles of carousels. The difference between the title of the first carousel (25.77\%) and its first item (90.82\%) reveals that at least 65.05\% of users examined the item but not the title.
  }
  \label{fig:exam-prob}
\end{figure*}

\section{An Analysis of User Examination Patterns}\label{sec:exp 1+2}

\subsection{The F-pattern does not generalize}
\label{sec:exp 1}

We start by considering Assumption~\ref{exp:expectation 1}: User examination follows an F-pattern where positions in the top-left region of the interface receive significantly more attention than others, with attention decaying towards the bottom and right.
The first assumption implies that the examination probability should be maximized at the top-left positions and decrease along the vertical and horizontal axes.
From the dataset, we can estimate the examination probability at position $(i, j)$ with the observed frequency of examination: %
\begin{equation}
P\!\left(E_{i,j}=1\right) \approx
\hat{P}\!\left(E_{i,j}=1\right) \coloneq \frac{N(E_{i,j}=1)}{N_{\text{total}}},
\label{eq:exam-prob}
\end{equation}
where $N(\cdot)$ denotes the number of screens that satisfy the condition and $N_{\text{total}}$ the total number of screens.
The resulting examination frequencies are displayed in Figure~\ref{fig:exam-prob} as percentages in a heatmap, where higher probabilities correspond to a higher color density.

If Assumption~\ref{exp:expectation 1} is correct then the heatmap should display the highest density in the top-left corner, with a gradual decay both vertically (top to bottom) and horizontally (left to right).
However, the results in Figure~\ref{fig:exam-prob} only partially follow this pattern.
On the first page, we see that the highest density is in the top-left corner and that, for the most part,  examination decays horizontally and vertically as assumed.
But closer inspection reveals another peak of examination on the top-right corner of Page~1, indicating a dual-focus structure that contradicts Assumption~\ref{exp:expectation 1}.
Moreover, when we consider the examination frequencies on the second and third page, we see that a horizontal decay is present but in the opposite direction (right to left).
Thus, revealing two mirrored F-shaped patterns \emph{within} Page~2 and~3.
If we consider all three pages at once, we see that there is no gradual change when crossing between pages, evidently there is \emph{no F-pattern that goes across pages}.
This strongly refutes Assumption~\ref{exp:expectation 1} on this dataset.

Potentially, the patterns we observe can be partially attributed to the interactive mechanics of carousel interfaces.
In this case, users have to click on an arrow on the right side of the screen to move a carousel from one page to another (swiping).
As a result, the users' examination will start on the right side of the second and third page after a swipe is made.
It is possible that the second peak of attention on the top-right of Page~1 is due to users looking for the swipe button, resulting in an increase in examination in that area.
However, we do not observe a similar consistent increase across the entire right column of Page~1, thus it seems unlikely that this is the correct explanation for the second peak.

In summary, we conclude that \emph{the examination data disproves Assumption~\ref{exp:expectation 1}: in carousel interfaces, user attention does \emph{not} follow a consistent F-pattern.}
Instead, we observe a dual-focus F-shaped pattern on the first page that shifts to mirrored F-shaped patterns within subsequent pages after swiping.

\begin{figure*}[!ph]
  \centering
  \textbf{Empirical conditional examination frequency}
  \includegraphics[clip, trim=0mm 317mm 0mm 10mm, width=\linewidth]{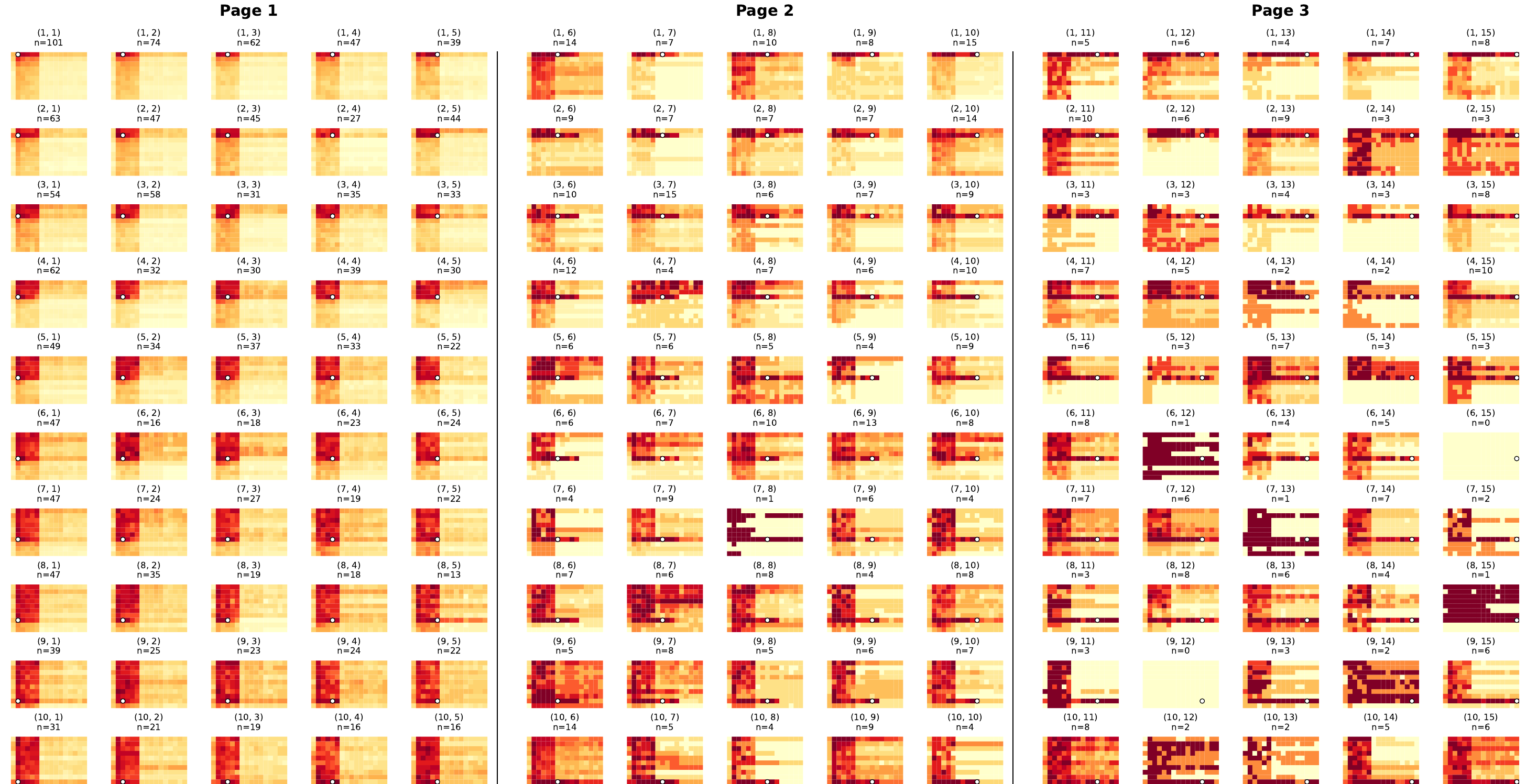}
  \textbf{Machine learning-smoothed conditional examination frequency}
  \includegraphics[clip, trim=0mm 0mm 0mm 307mm, width=\linewidth]{figures/combined_raw_and_smoothed.pdf}
  \caption{
  In context of evaluating Assumption~\ref{exp:expectation 2}: The raw and machine-learning–smoothed examination frequency, conditioned on a click at a specific position.
  Each subplot corresponds to one conditioning position (highlighted by a white dot and labeled above); for the empirical heatmap, the number of observed clicks at that position ($n$) is additionally annotated. 
  Within a subplot, each cell represents the raw or smoothed frequency that the corresponding position is examined prior to the click, given that the final click occurs at the highlighted position.
  Overall, both heatmaps illustrate L-shaped patterns in which the positions are almost certainly examined before being clicked.
  }
  \label{fig:raw-heatmap-before-click}
  \label{fig:conditioned-examination}
\end{figure*}

\subsection{Examination conditioned on clicks is not hierarchical but follows an L-shape}\label{sec:exp 2}
We continue with Assumption~\ref{exp:expectation 2}: user examination conditioned on a click follows an hierarchical pattern where items to the left of the clicked position \emph{are} examined, items to the right and in subsequent carousels (below) \emph{are not} examined, and items in preceding carousels (above) \emph{could} be examined.
This assumption corresponds to a sequential and hierarchical click model where users consider the carousels in order (top to bottom) and decide per carousel whether to enter it and examine the items within, which are also examined in order (left to right).

\begin{figure*}[t]
\centering
\includegraphics[width=\linewidth, alt={Netflix carousel interface}]{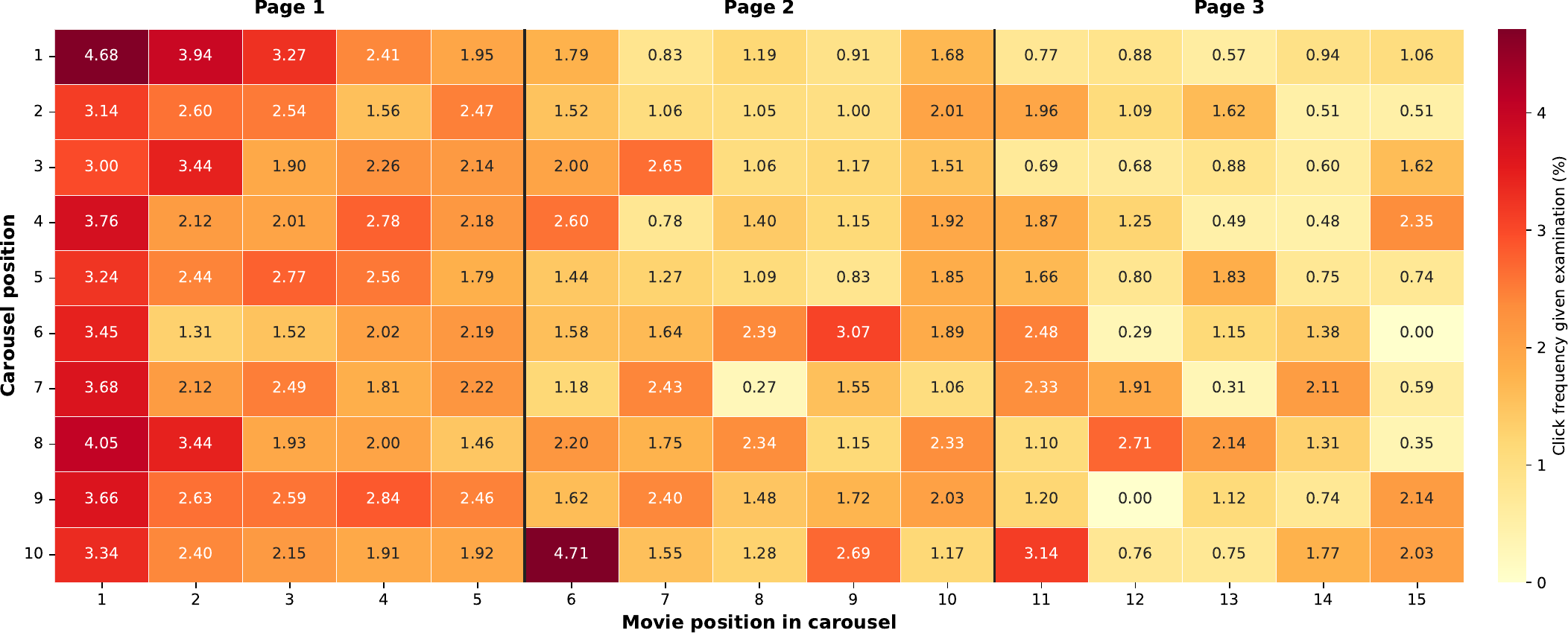}
\caption{ Click frequency given examination per item position in percentage. The positions $(6,15)$ and $(9,12)$ are never clicked in the dataset, so the conditional click frequency is $0$.
Refuting Assumption~\ref{exp:expectation 3}: There are significant vertical fluctuations (within columns) which should not occur under the examination hypothesis due to the randomization of the carousel order.}
\label{fig:ctr}
\end{figure*}

To verify whether the assumed pattern occurs in the dataset, we need an empirical estimate of the conditional probability of examination of position $(i, j)$ given a click on position $(i', j')$.
We use the following observed conditional frequency:
\begin{equation} 
\hat{P}\left(E_{i,j}=1 \mid C_{i',j'}=1\right) \coloneq \frac{N(E_{i,j}=1 \land C_{i',j'}=1)}{N(C_{i',j'}=1)}. 
\end{equation}
A difficulty with this empirical estimate is that the condition may seldom occur, resulting in a very low $N(C_{i',j'}=1)$ and thus a very high variance.
Moreover, the positions $(6,15)$ and $(9,12)$ are never clicked in the dataset (e.g., $N(C_{6,15}=1) = 0$) resulting in undefined estimates.
As a solution, we apply machine-learning-based smoothing as discussed in Section~\ref{sec: data smoothing}.

To assess how closely the smoothed predictions track the empirical data, we compute two complementary metrics between smoothed and raw 
frequencies across positions with sufficient click data: the Pearson 
correlation, which captures whether the two align in shape, and the 
mean absolute error (MAE), which captures how far they differ in 
magnitude.
Restricting the comparison to the 63 positions with at least 10 observed clicks, we find that the smoothed predictions align closely with the raw frequencies ($r = 0.95$, MAE~$= 6.05$~pp), indicating that the model faithfully preserves the structure of the data in regions where the raw frequencies are reliable.
At positions with sparser observations, however, the smoothed estimates rely more heavily on interpolation from neighboring positions, and the spatial smoothness assumption implies that if a position truly deviates from its neighbors, the smoothed estimate will partially flatten such deviations toward the surrounding pattern.

All conditional estimates are shown in Figure~\ref{fig:conditioned-examination}, which displays two examination heatmaps (empirical and smoothed) for every possible click position. 
We see that the smoothing makes a substantial difference on Pages~2 and~3, where clicks are considerably sparser than on Page~1.

According to Assumption~\ref{exp:expectation 2} the click position should act as an obvious visual boundary: the heatmap should display a high density to the left of the click position, whereas the regions to the right and below it should remain largely empty.
The assumption says nothing about the carousels above the click position, which could thus display any density.
However, when we inspect Figure~\ref{fig:conditioned-examination}, we see that the implications of the assumption do not bear out:
We observe that clicks on Pages~2 and~3 co-occur with high frequencies of examination on the entire first page; furthermore, for clicks that are \emph{not} on the right edge of a page (columns 1--4, 6--9 and 11--14), it is nearly certain that items to their right are also examined.
Both of these observations contradict the implications of Assumption~\ref{exp:expectation 2}, and therefore, provide strong evidence \emph{against} it.

Instead of the assumed examination pattern, we instead observe a different but very clear pattern:
On the first page, all items in carousels above the clicked position are \emph{nearly universally} examined, additionally, items in the clicked carousel that are on the same page as the click or on preceding pages are also \emph{almost always} examined.
We name this phenomenon the \emph{L-pattern} as the first page serves as a thick vertical line that ends in a thin horizontal line that continues until the end of the clicked page, resembling the letter \emph{L}.
In addition to the L-pattern, we see that items on the first page in carousels below the click position also have a high probability of examination.
To a lesser extent, items on subsequent pages in carousels above the click position also have an increased examination probability.

To summarize our observations; examination conditioned on clicks shows a clear L shape of items that are almost certainly examined, we propose to call this the \emph{L-pattern}.
Together with the L-pattern, we also see an increased probability on the remainder of the first page, and items on subsequent pages above the clicked carousel.
These examination patterns contradict those implied by Assumption~\ref{exp:expectation 2}, and therefore, we confidently \emph{reject} it for this dataset.

\section{Refuting the Examination Hypothesis}
\label{sec:exp 3}
According to Assumption~\ref{exp:expectation 3}, the examination hypothesis holds for carousel interfaces. This means that the conditional click probability of an item given examination does not vary over display positions.
In other words, let $I_{i,j}$ denote the item at position $(i,j)$, then for all $x$, $(i,j)$, $(i',j')$:
\begin{align}
P(C_{i,j} = 1 \mid E_{i,j} = 1, I_{i,j}=x)
= P(C_{i'\!,j'\!} = 1 \mid E_{i'\!,j'\!} = 1, I_{i'\!,j'\!}=x).
\nonumber
\end{align}

At first glance, it seems natural to estimate the conditional click probability using the following empirical frequency:
\begin{equation} 
\hat{P}(C_{i,j} = 1 \mid E_{i,j} = 1, I_{i,j}=x) \coloneq \frac{N(E_{i,j}=1 \land C_{i,j}=1 \land I_{i,j} = x)}{N(E_{i,j}=1 \land I_{i,j} = x)},
\end{equation}
and subsequently compare these estimates for the identical item but differing positions.
Unfortunately, the dataset does not allow for such a comparison since each item is displayed at the same position to every user, thus, we lack any second estimates to compare with.
However, the (vertical) order of the carousels are completely randomized per screen, thus over the dataset, there is statistically an identical distribution of items over each position in the same column.
Therefore, Assumption~\ref{exp:expectation 3} implies that there is no difference between conditional click probabilities in the same column \emph{given} examination but \emph{not given} an item.
Formally,
\begin{align}
\forall (i,j), i' \;;\;
P(C_{i,j} = 1 \mid E_{i,j} = 1)
= P(C_{i'\!,j} = 1 \mid E_{i'\!,j} = 1).
\nonumber
\end{align}
We note that there is no similar implication between positions in the same row, since item ordering within carousels is not randomized but ordered by popularity.
We estimate the conditional click probability given examination as follows:
\begin{equation} 
\hat{P}\left(C_{i,j}=1 \mid E_{i,j}=1\right) \coloneq \frac{N(E_{i,j}=1 \land C_{i,j}=1)}{N(E_{i,j}=1)}. 
\end{equation}

The estimates are shown as percentages in the heatmap in Figure~\ref{fig:ctr}.
Thus, if Assumption~\ref{exp:expectation 3} holds then the heatmap should show similar conditional click frequency estimates across positions in a column, due to the randomization of carousel order.
However, we actually observe that the conditional click probability shows substantial fluctuations within columns, particularly on Page 1.
This appears to reveal a significant deviation from the examination hypothesis as it indicates that the conditional click probability is actually statistically dependent on display position.
To rigorously verify this difference, we conducted five two-sided Z-tests to compare the conditional frequencies in Row 1 (first carousel) against Row 6 (sixth carousel) in the first five columns (Page~1), the results are displayed in Table~\ref{tab:ctr_comparison_final}.
This statistical analysis confirms significant differences in Column~2 and~3, where the conditional click probability is significantly higher in Row 1 than in Row 6 with $p < 0.05$.
Although we do not find statistically significant differences for the other columns, the significant differences found are sufficient to refute Assumption~\ref{exp:expectation 3} and thus the examination hypothesis as well.

To summarize, we have analyzed the conditional click frequency per position given examination, due to the randomization of carousel orderings these frequencies should be equal within columns when the examination hypothesis is true.
However, we found statistically significant differences within columns, meaning that we can \emph{reject} Assumption~\ref{exp:expectation 3}, and by extension, we can \emph{refute} the examination hypothesis on this dataset as well.

\begin{table}[t]
  \centering
  \setlength{\tabcolsep}{2pt}
  \caption{
  Clicks and estimation frequencies for Row 1 vs.\ Row 6 and percentage differences between the click frequencies conditioned on examination, along with p-values computed for these differences.
  The absolute conditional click frequencies are reported in Figure \ref{fig:ctr}.
  Differences are only measured within columns. 
  Statistical significance is assessed using two-sided Z-tests, with significance at $p < 0.05$ ($p < 0.01$  after Bonferroni correction) indicated in bold. }
  \label{tab:ctr_comparison_final}
  \small  
  \begin{threeparttable}
    \resizebox{\linewidth}{!}{
        \begin{tabular}{c cc cc cc}
        \toprule
        \textbf{Col.} & \multicolumn{2}{c}{\textbf{Row 1}} & \multicolumn{2}{c}{\textbf{Row 6}} & \textbf{Diff.} & \textbf{\textit{p}-value} \\
        \cmidrule(lr){2-3} \cmidrule(lr){4-5}
         & Click & Examination & Click & Examination &  &  \\
        \midrule
        
        1 & 101 & 2157 & 47 & 1363 & $+$1.23\% & 0.076 \\
        
        2 & \phantom{0}\textbf{74} & \textbf{1880} & \textbf{16} & \textbf{1226} & \textbf{$+$2.63\%} & \llap{$<$}\textbf{0.001}\\
    
        3 & \phantom{0}\textbf{62} & \textbf{1895} & \textbf{18} & \textbf{1183} & \textbf{$+$1.75\%} & \textbf{0.003}\\
        
        4 & \phantom{0}47 & 1952 & 23 & 1141 & $+$0.39\% & 0.479 \\
        
        5 & \phantom{0}39 & 1995 & 24 & 1097 & $-$0.24\% & 0.661 \\
        
        \bottomrule
        \end{tabular}
    }
  \end{threeparttable}
\end{table}

\begin{figure*}[t]
\centering
\includegraphics[width=\linewidth, alt={Netflix carousel interface}]{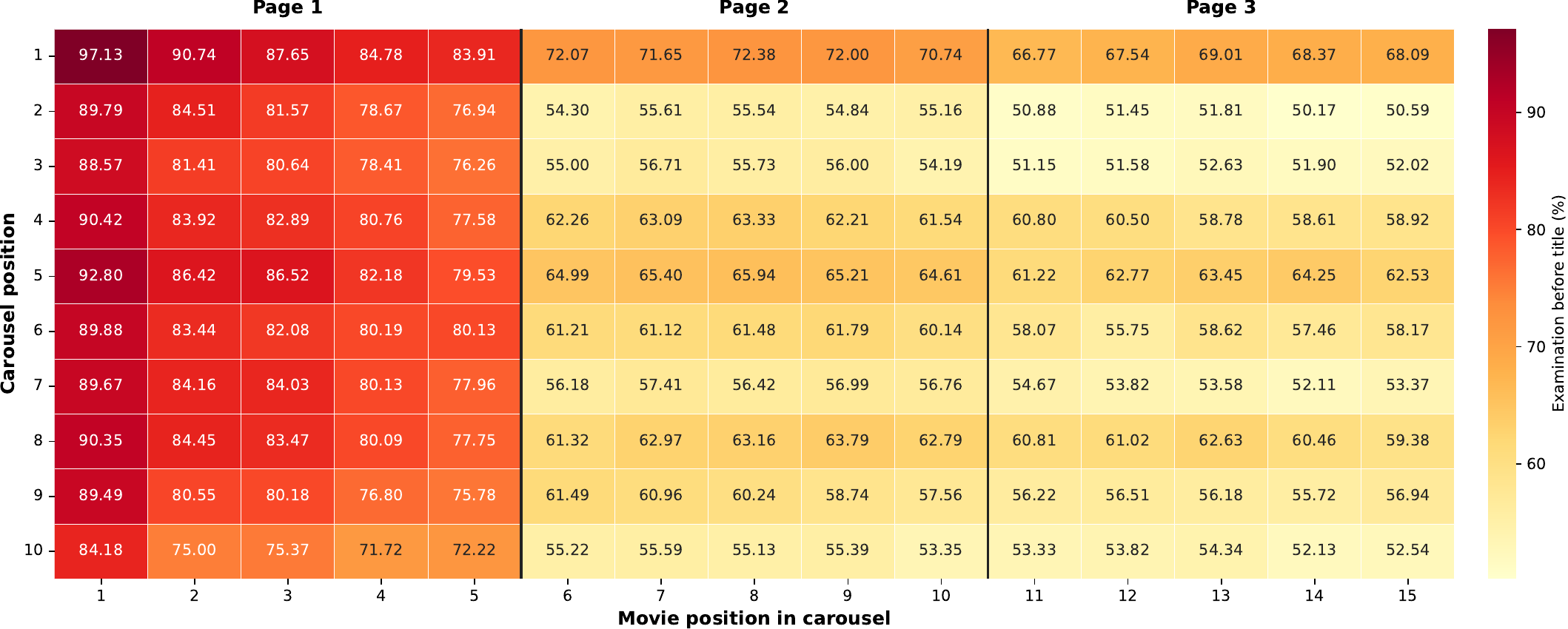}
\vspace{-1.5\baselineskip}
\caption{Empirical frequency of item positions being examined earlier than their corresponding title, conditioned on examination of the item positions. Refuting Assumption~\ref{exp:expectation 4}: All frequencies are over 50\%, meaning items are ubiquitously examined first.}
\label{fig:first arrival}
\end{figure*}

\section{Most Users Ignore Carousel Headers}
\label{sec:exp 4}
Lastly, we turn to the Assumption~\ref{exp:expectation 4} which states that users examine the title of the carousel before examining any individual items within it.
This assumption stems from the seemingly reasonable expectation that carousel titles serve as a navigation tool and thus should have a large influence on user behavior.
Moreover, titles are often specifically designed so that users can use them to decide whether the items inside are worth considering.
Therefore, this assumption implies that the titles of carousels should not only receive more examinations, but also receive earlier examination than their contents.

We begin our analysis by reconsidering Figure~\ref{fig:exam-prob}, which displays the examination frequencies of all item positions but also the title positions.
If carousel titles are examined before their items, then for each row, the title examination frequency should be higher than for all the items within the row.
Surprisingly, this is clearly not the case, we see that every row has several item positions that are examined considerably more frequently than their corresponding title positions.
The most extreme example is the left-most item position of the top row which is examined by 90.82\% of users, whereas the title of the row is examined by only 25.77\% of users.
Therefore, at least 65.05\% of users examined the left-top item but not the top carousel title.
Moreover, Figure~\ref{fig:exam-prob} reveals that across all rows of the left-most item column positions are examined at least twice as often as their accompanying carousel titles.
Clearly, we can conclude that Assumption~\ref{exp:expectation 4} \emph{does not hold} for this dataset.

Although Assumption~\ref{exp:expectation 4} is refuted, we want to evaluate to what extent it is accurate; 
We conduct a deeper analysis by addressing two further questions: (i) How frequent are titles the first element examined in a carousel? and (ii) Which item positions are frequently examined earlier than their corresponding carousel titles?

To answer the first question, we consider the empirical frequency that the title is examined before any of the its carousel items, given that at least one of the items or the title is examined (or both):
\begin{equation}
\begin{split}
&\hat{P}\left(\forall j > 0 \;; \; E_{i,0} \prec E_{i,j} \mid \exists j \geq 0; E_{i,j} = 1) \right)\\
&\qquad\qquad\qquad\coloneq \frac{N(\exists j \geq 0; E_{i,j} = 1 \land \forall j >0 ;  E_{i,0} \prec E_{i,j})}{N(\exists j \geq 0; E_{i,j} = 1)}.
\end{split}
\end{equation}
In cases where a title was examined but none of its carousel items, we consider the title to be examined earlier and vice-versa when the title is not examined but any of the items are.
The estimated frequencies for carousels in rows 1 through 10 are 
[2.06\%, 6.60\%, 9.27\%, 8.13\%, 5.75\%, 8.11\%, 8.52\%, 8.53\%, 9.00\%, 15.09\%], 
indicating that the title is rarely the initial point of attention. 
This finding reinforces our rejection of Assumption~\ref{exp:expectation 4}, as we see that, regardless of the row, users seldom examine the carousel title before its items.
This holds even for the last row which has the highest frequency with 15.09\%, as this still means that on the large majority of screens users did not examine the title first.

For the second question, we compute the empirical frequency of individual item positions being examined earlier than their corresponding title, conditioned on the item positions being examined:
\begin{equation}
\hat{P}\left(E_{i,0} \succ E_{i,j} \mid E_{i,j}=1 \right)
\coloneq \frac{N(E_{i,0} \succ E_{i,j} \land E_{i,j}=1)}{N(E_{i,j}=1)}. 
\end{equation}
The results are presented as a heat map in Figure~\ref{fig:first arrival}.
The figure reveals that positions on the leftmost column have frequencies of over 88\%, with the sole exception of the last carousel, which remains high at approximately 84\%. 
Across the entire first page, we see that most positions are observed before their title in 80\% of cases.
The frequencies are lower on subsequent pages, however, all items positions have a frequency of over 50\% of being examined before their carousel title; clearly refuting Assumption~\ref{exp:expectation 4}.

In summary, our findings demonstrate that carousel titles are not only infrequently examined, but that even when observed, they are typically still examined after the items within their carousel. 
Since the empirical evidence strongly contradicts the primacy of carousel titles, we confidently reject Assumption~\ref{exp:expectation 4} for this dataset.

\section{Discussion and Implications}
Our analysis of the RecGaze dataset has provided several key insights into user behavior within carousel interfaces, which have contradicted all of the assumptions that Section~\ref{sec:expect} posited based on previous work.
Instead we made several surprising new observations:
For instance, Figure~\ref{fig:raw-heatmap-before-click} suggests that users rarely skip carousels but instead tend to systematically examine all visible items on the first page of preceding carousels.
Horizontal swipes mostly only take place in the carousel that receives the final click, and user examination usually does not go beyond the page of the clicked item.
Furthermore, movie thumbnails consistently attract more user examinations than carousel titles (Figure~\ref{fig:exam-prob}) and users typically examine the content of carousels before the corresponding titles, and do not appear to use titles as navigation tools (Figure~\ref{fig:first arrival}).
Furthermore, Figure~\ref{fig:ctr} shows that the conditional click probability of an item, given examination, varies between different rows within the same column.
This phenomenon appears very similar to \emph{trust bias} in single-list interfaces, where users click top-ranked results more often due to a greater perceived trust in the ranking order~\cite{10.1145/3308558.3313697, 10.1145/3340531.3412031}. 

Our observations have implications for the design of evaluation metrics tailored to carousel-based interfaces.
For example, traditional evaluation metrics typically employ a monotonically decreasing discount factor that starts in the top-left corner of the interface and decreases in horizontal and vertical directions.
In contrast, our results demonstrate that such discounting does not correspond to user examination patterns in carousel interfaces.
Instead, discounting should follow a dual-focus pattern on the first carousel page and a mirrored F-pattern on subsequent pages, corresponding to those indicated by the examination data (Figure~\ref{fig:exam-prob}).

Our findings also have implications on click model design for carousel settings.
Importantly, we find strong evidence that the examination hypothesis, a foundational assumption in most click models, does not hold in carousel settings.
This invalidation highlights the importance of carefully selecting what variables are modelled to influence click decisions, and the difficulty of disentangling presentation and preference factors.
Moreover, we find that visual information, such as movie thumbnails, not only attracts more examination than carousel titles but also captures it significantly earlier.
This finding strongly rejects the concept of hierarchical click models that assume the carousel titles are used to decide whether their contents should be examined~\citep{10.1145/3643709}.
Whilst introducing novel click models is out of the scope of this work, our empirical study does rule out several classes of potential click model designs.

\section{Conclusions}
In this paper, we performed a comprehensive analysis of the examination and clicking behavior of users within carousel interfaces, using the RecGaze eye-tracking dataset.
The code for our analysis is available at 
\url{https://github.com/jkang98/RecGaze-Analysis}.

We started by identifying several assumptions derived from earlier work on web-search interfaces, which have been applied to the carousel setting by previous literature.
Subsequently, the findings of our analysis have refuted all of these behavioral assumptions, highlighting that user behaviors in carousel interfaces differ fundamentally from traditional web search.
Additionally, we have observed several behaviors that have not been recognized by earlier work, such as the L-pattern in click-conditioned examination frequencies.
Based on these new insights, we have articulated specific implications for the design of evaluation metrics and click models, laying a solid foundation for future research.

Our work also has several limitations that point to directions for future research.
First, our analysis is based on the RecGaze dataset, which to our knowledge is the only publicly available eye-tracking dataset for carousel interfaces. 
While this limits the diversity of carousel UI designs we can study, RecGaze does contain other tasks beyond the free-browsing setting considered here. 
Since UI design and task framing are known to influence user behavior, our conclusions may not generalize to other carousel layouts or interaction tasks. Extending our analysis to the other tasks in RecGaze, as well as to new datasets collected under different carousel layouts, is an important direction for future work.
Second, we treat users as a homogeneous group and do not analyze how attention patterns vary across individuals. Given that browsing behavior likely differs substantially from person to person, examining such individual differences is important for future work.
Finally, the analysis of this work only considered fixation and clicks, leaving the analysis of additional interaction signals, such as cursor movement trajectories, and features regarding users' decision-making, such as movie familiarity and self-reported reasons for clicking, for future exploration.

\begin{acks}
This work is supported by the Dutch Research Council (NWO) under grants \href{https://www.nwo.nl/en/projects/viveni222269}{VI.Veni.222.269}, \href{https://www.nwo.nl/en/projects/024004022}{024.004.022}, \href{https://www.nwo.nl/en/projects/nwa138920183}{NWA.1389.20.183}, and \href{https://www.nwo.nl/en/projects/kich3ltp20006}{KICH3.LTP.20.006}.
It is also supported by the European Union under grant agreement No.\ \href{https://doi.org/10.3030/101201510}{101201510}.
All content represents the opinion of the authors, which is not necessarily shared or endorsed by their respective employers and/or sponsors.
\end{acks}

\bibliographystyle{ACM-Reference-Format}
\balance
\bibliography{references.bib}

\end{document}